\documentclass[Physsubmission, Phys]{SciPost}

\usepackage{amsmath}
\usepackage{amsfonts}
\newcommand{\jpsi}{$\mathrm{J}/\psi$}
\newcommand{\jpsim}{\mathrm{J}/\psi}
\newcommand{\vect}[1]{\boldsymbol{#1}_{\perp}}

\usepackage[caption = false]{subfig}

\newcommand{\xt}{{\vect{x}}}
\newcommand{\bt}{\vect{b}}

\newcommand{\yt}{{\vect{y}}}
\newcommand{\rt}{\vect{r}}

\newcommand{\xpom}{{x_\mathbb{P}}}

\newcommand{\gev}{\ \textrm{GeV}}

\newcommand{\Mgammastar}[1]{\widetilde{\mathcal{M}}_{#1}^{\gamma^* A,\gamma^*}}

\newcommand{\der}{\mathrm{d}}

\newcommand{\Deltat}{\mathbf{\Delta}}

\newcommand{\phik}{\phi_{k\Delta}}
\newcommand{\phirb}{\phi_{r b}}

\hypersetup{
    colorlinks,
    linkcolor={red!50!black},
    citecolor={blue!50!black},
    urlcolor={blue!80!black}
}

\DeclareSymbolFont{usualmathcal}{OMS}{cmsy}{m}{n}
\DeclareSymbolFontAlphabet{\mathcal}{usualmathcal}

\begin{document}

\begin{center}{\Large \textbf{
Azimuthal correlations in diffractive scattering at the Electron-Ion Collider\\
}}\end{center}

\begin{center}
A. Dumitru\textsuperscript{1,2},
H. Mäntysaari\textsuperscript{3,4$\star$},
R. Paatelainen\textsuperscript{4},
K. Roy\textsuperscript{5,6},
F. Salazar\textsuperscript{6,7,8},
B. Schenke\textsuperscript{7}
\end{center}

\begin{center}
{\bf 1} Department of Natural Sciences, Baruch College, CUNY, 17 Lexington Avenue, New York, NY 10010, USA
\\
{\bf 2} The Graduate School and University Center, The City University of New York, 365 Fifth Avenue, New York, NY 10016, USA
\\
{\bf 3} Department of Physics, University of Jyväskylä,  P.O. Box 35, 40014 University of Jyväskylä, Finland
\\
{\bf 4} Helsinki Institute of Physics, P.O. Box 64, 00014 University of Helsinki, Finland
\\
{\bf 5} Max-Planck-Institut f\"{u}r Physik, F\"{o}hringer Ring 6, 80805 M\"{u}nchen, Germany \\
{\bf 6} Department of Physics and Astronomy, Stony Brook University, Stony Brook, NY 11794, USA\\
{\bf 7} Physics Department, Brookhaven National Laboratory, Bldg. 510A, Upton, NY 11973, USA\\
{\bf 8} Center for Frontiers in Nuclear Science (CFNS), Stony Brook University,
Stony Brook, NY 11794, USA

* heikki.mantysaari@jyu.fi
\end{center}

\begin{center}
\today
\end{center}


\definecolor{palegray}{gray}{0.95}
\begin{center}
\colorbox{palegray}{
  \begin{tabular}{rr}
  \begin{minipage}{0.1\textwidth}
    \includegraphics[width=22mm]{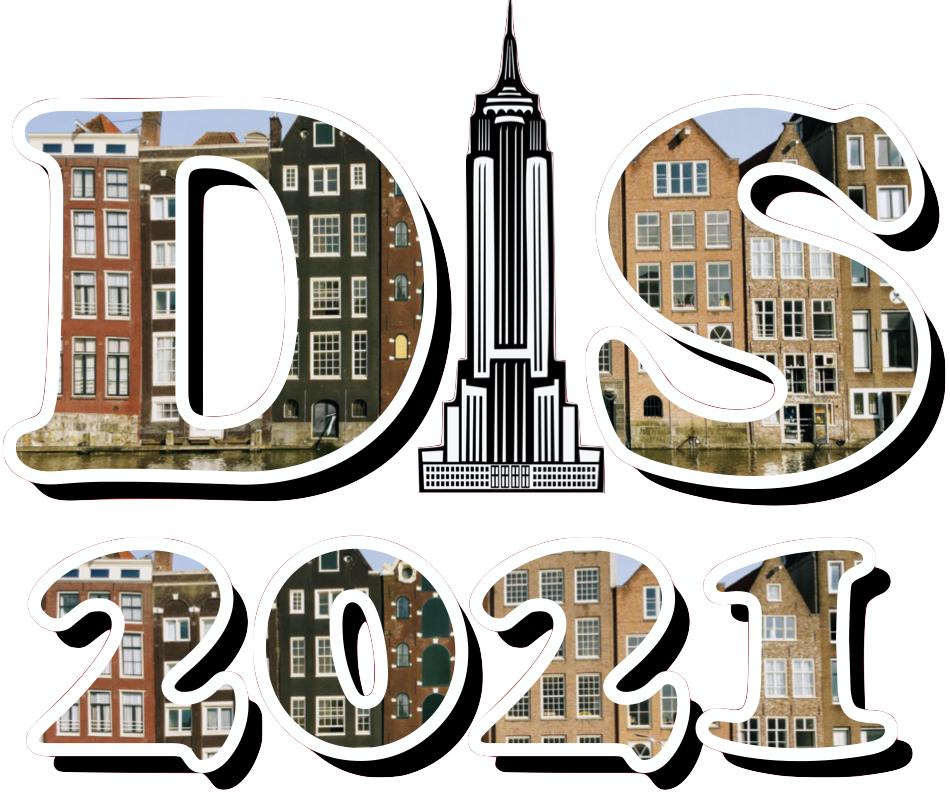}
  \end{minipage}
  &
  \begin{minipage}{0.75\textwidth}
    \begin{center}
    {\it Proceedings for the XXVIII International Workshop\\ on Deep-Inelastic Scattering and
Related Subjects,}\\
    {\it Stony Brook University, New York, USA, 12-16 April 2021} \\
    \doi{10.21468/SciPostPhysProc.?}\\
    \end{center}
  \end{minipage}
\end{tabular}
}
\end{center}

\section*{Abstract}
{\bf
We calculate azimuthal correlations between the exclusively produced vector meson and the scattered electron in Deep Inelastic Scattering processes at the future Electron-Ion Collider (EIC). We identify ``kinematical'' and ``intrinsic'' contributions to these correlations, and show that the correlations are sensitive to the non-trivial correlations in the gluon distribution of the target. Realistic predictions at the EIC kinematics are provided using two different approaches to describe the dipole-proton interaction at relatively small $x$.
}


\section{Introduction}
\label{sec:intro}
Exclusive vector particle production at high energy is a powerful probe of the small-$x$ structure of protons and nuclei. This process has two particular advantages. First, as there is no net color charge transfer to the target, the cross section scales approximately as the gluon density \emph{squared}~\cite{Ryskin:1992ui}. Additionally, as the momentum transfer is Fourier conjugate to the impact parameter, it becomes possible to study the spatial distribution of small-$x$ gluons (and their event-by-event fluctuations)~\cite{Mantysaari:2016ykx,Mantysaari:2016jaz,Mantysaari:2020axf}.

To perform more detailed imaging, it is useful to look at more differential measurements. In Ref.~\cite{Mantysaari:2020lhf} presented in this Talk, exclusive \jpsi\ and photon production cross sections have been calculated differentially in both squared momentum transfer $|t|$ and the azimuthal angle $\phi_{k\Delta}$ between the outgoing electron and the produced vector particle in the frame where the incoming photon has no transverse momentum.

\section{Azimuthal correlations in exclusive cross section}

Cross section for exclusive vector particle production in electron-hadron collisions was calculated in Ref.~\cite{Mantysaari:2020lhf}. The cross section can be written as
\begin{multline}
   \frac{\mathrm{d} \sigma^{e A \rightarrow e A V}}{\der \xpom \der Q^2 \der |t| \der \phi_{k\Delta}} = \frac{\alpha_{em}}{32 \pi^3 Q^2 \xpom} \sum_{\lambda'=0,\pm 1}  \left \{ (1-y)  \left|\left\langle \widetilde{\mathcal{M}}^{\gamma^*A,V}_{0,\lambda'}\right \rangle_Y \right|^2 \right. \\
\left.   + \frac{1}{4} \left[ 1 + (1-y)^2 \right] \sum_{\lambda=\pm 1} \left| \left \langle \widetilde{\mathcal{M}}^{\gamma^*A,V}_{\lambda,\lambda'} \right \rangle_Y \right|^2 \right.  \\
     - \frac{\sqrt{2}}{2} (2-y) \sqrt{1-y} \sum_{\lambda=\pm 1} \mathrm{Re}  \left(\left\langle\widetilde{\mathcal{M}}^{\gamma^*A,V}_{0,\lambda'} \right \rangle_Y \left \langle \widetilde{\mathcal{M}}^{\gamma^*A,V}_{\lambda,\lambda'} \right \rangle_Y^*  \right) \cos( \phi_{k\Delta})  \\  
     + \left. (1-y) \mathrm{Re} \left( \left \langle \widetilde{\mathcal{M}}^{\gamma^*A,V}_{+1,\lambda'} \right \rangle_Y  \left \langle \widetilde{\mathcal{M}}^{\gamma^*A,V}_{-1,\lambda'} \right \rangle_Y^* \right) \cos(2 \phi_{k\Delta})  \right \}.
    \label{eq:xs}
\end{multline}
Here $\mathcal{M}^{\gamma^*A,V}_{\lambda,\lambda'}$ corresponds to the scattering amplitude for the $\gamma^* + A \to V + A$ scattering, where $\lambda$ and $\lambda'$ refer to the polarizations of the $\gamma^*$ and $V$, respectively,  $V=\jpsim$ or $V=\gamma$ and $y$ is the inelasticity variable. Explicit expression for these amplitudes are calculated in Ref.~\cite{Mantysaari:2020lhf} and the results are in agreement with the previous calculation performed in the same Color Glass Condensate (CGC) framework~\cite{Hatta:2017cte}. 

An important feature in Eq.~\eqref{eq:xs} is that the interference between amplitudes with different polarization states results in non-zero $\cos(\phik)$ and $\cos(2\phik)$ modulations. These are in turn related to the dependence on orientation between the impact parameter $\bt$ and dipole size $\rt$ in the dipole-target scattering amplitude $\mathcal{D}(\rt,\bt)$. For example, the amplitude $\Mgammastar{\pm 1, \mp 1}$ reads
\begin{multline}
    \Mgammastar{\pm 1, \mp 1} = -\frac{8 N_c e^2 q_f^2}{(2\pi)^2}  \int_{\vect{b}}  e^{-i \vect{\Delta} \cdot \bt } \int_{\vect{r}} e^{\pm 2 i  \phi_{r \Delta}} \mathcal{D}(\rt,\bt) \\
    \times \int_{z} e^{-i  \vect{\delta} \cdot \vect{r}} z \bar{z} \ \varepsilon_f  K_1(\varepsilon_f  r_\perp) \varepsilon_f'  K_1(\varepsilon_f'  r_\perp) \, , 
    \label{eq:M+-}
\end{multline}
where $\varepsilon_f^2 = Q^2z\bar z+ m_f^2 $ with $\bar z=1-z$. As $\phi_{r \Delta}$ is the azimuthal angle between the dipole orientation and the momentum transfer vectors, this amplitude is directly sensitive to the $\cos (2\phirb)$ modulations in the dipole amplitude, where $\phirb$ is the angle between the impact parameter and the dipole orientation. This contribution to the azimuthal modulations in Eq.~\eqref{eq:xs} we refer to as the intrinsic contribution. There is also an additional purely kinematical contribution from the non-zero off-forward phase factor $ \vect{\delta} = \frac{z-\bar z}{2} \Deltat$.

\section{Numerical setup}
\label{sec:setup}
The main results in Ref.~\cite{Mantysaari:2020lhf} are obtained using a CGC setup, where the Wilson lines describing the target structure at initial $x=0.01$ are obtained from the McLerran-Venugopalan model~\cite{McLerran:1993ni,McLerran:1993ka} with impact parameter dependence. The non-uniform geometry is included by using a local Gaussian color charge correlator $g^2\langle \rho(\xt),\rho(\yt)\rangle \sim g^4\mu^2$, where the density $g^4\mu^2$ depends on the impact parameter $\bt=(\xt+\yt)/2$, and for the protons a Gaussian distribution $\sim e^{-\bt^2/(2B)}$ is assumed.
To calculate cross sections at $x<0.01$, this initial condition is coupled to a JIMWLK evolution equation~\cite{Mueller:2001uk}. 

The MV model with impact parameter dependence results in a dipole-proton scattering amplitude where the scattering probability is highest when the dipole is oriented parallel to the impact parameter (see~\cite{Mantysaari:2020lhf,Mantysaari:2019hkq,Iancu:2017fzn}). For comparison, in this Contribution we also present preliminary results using a recent dipole-target scattering amplitude obtained as a result of an explicit light cone perturbation theory calculation in Refs.~\cite{Dumitru:2021tvw,Dumitru:2020gla} (``LCPT dipole''). In that approach, one takes a non-perturbative wave function constrained by low-energy data to describe the proton structure in the valence quark region, and then calculates perturbatively a gluon emission which allows extension of the calculation towards moderately small $x$. This calculation results in non-trivial correlations that are absent in the MV model, and for example the dipole-target scattering amplitude is largest when the dipole is perpendicular to the impact parameter.

\section{Numerical results}

\begin{figure}[tb]
    \subfloat[Dependence on $\xpom$.]{
        \includegraphics[width=0.48\textwidth]{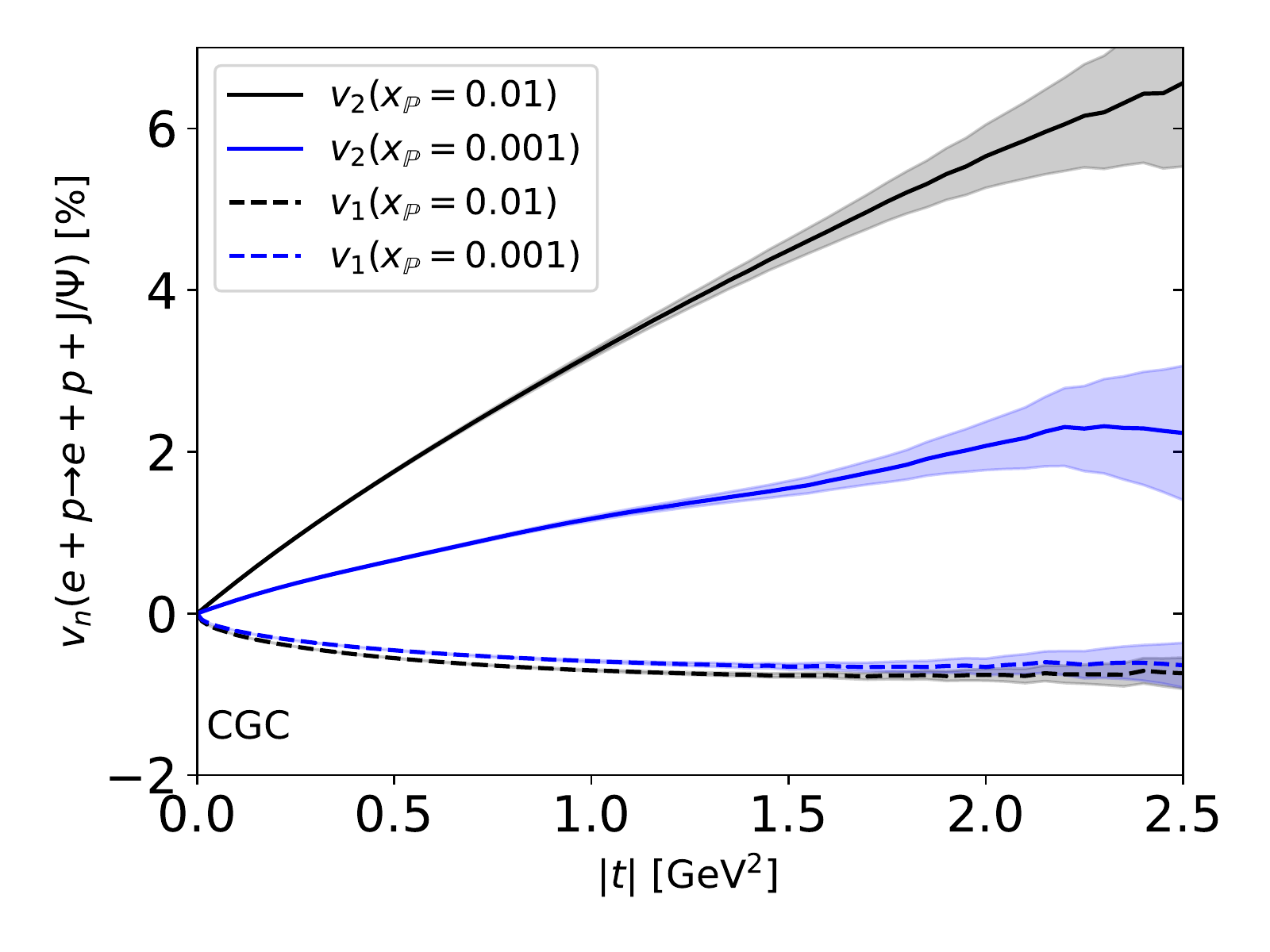}
        \label{fig:jpsi_vn_xdep}
        }
        \subfloat[MV model compared to the LCPT dipole.]
        {
        \includegraphics[width=0.48\textwidth]{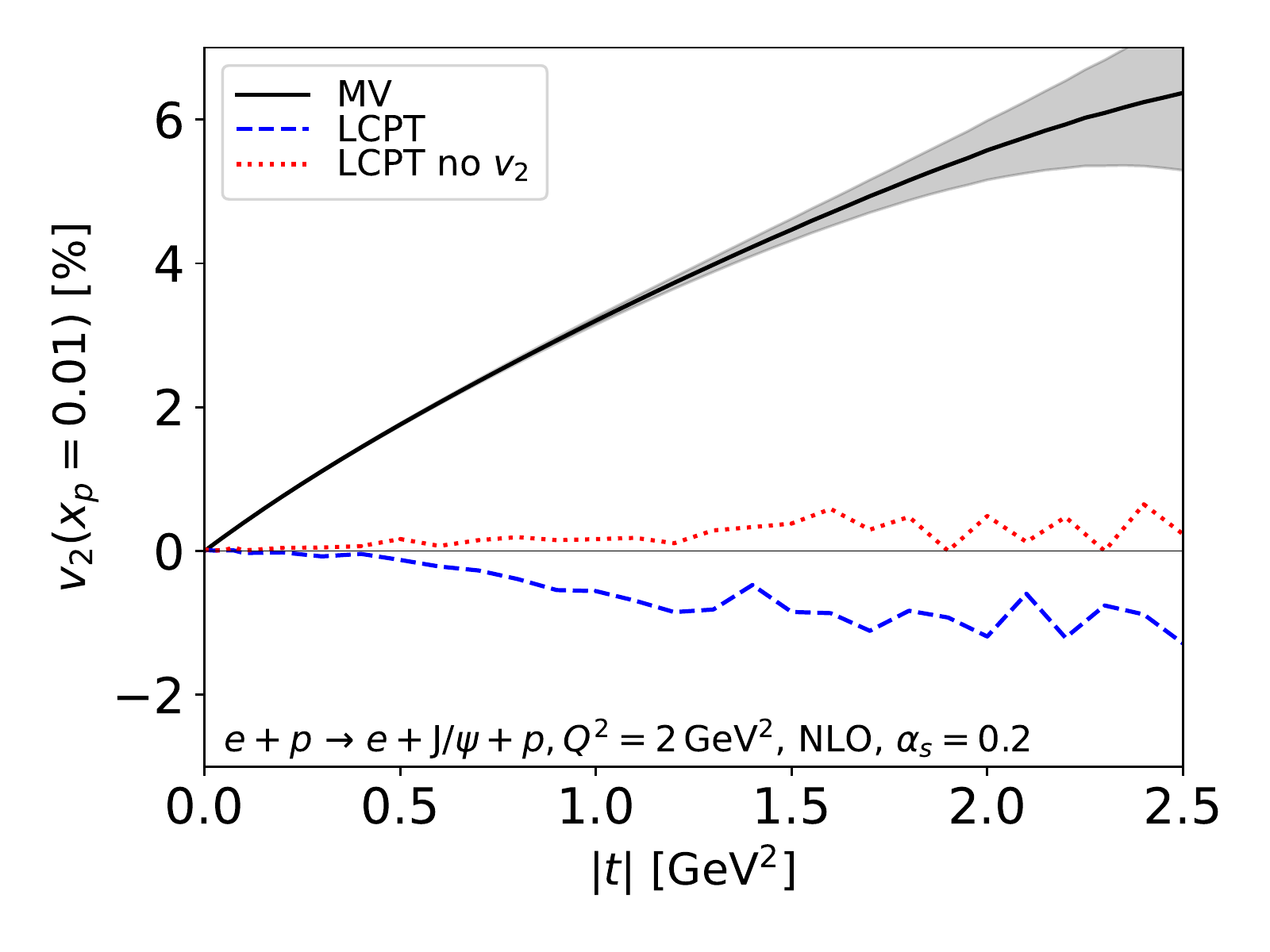}
         \label{fig:lcpt_v2}
        }
        \caption{The modulation coefficient $v_n$ in  coherent \jpsi\ production at $Q^2=2\gev^2$ from Ref.~\cite{Mantysaari:2020lhf}. Left: dependence on the momentum fraction $\xpom$. Right: $v_2$ calculated using the dipole amplitude from the impact parameter dependent MV model (standard setup) and from the LCPT calculation of Ref~\cite{Dumitru:2021tvw}.
        }
\end{figure}

In this Contribution we focus on exclusive \jpsi\ production in $e+p$ scattering, the results for DVCS are discussed in detail in Ref.~\cite{Mantysaari:2020lhf} as well as $e+A$ collisions where the angular modulations are found to be very small. We show results for the modulation coefficients $v_n$ defined as
\begin{equation}
\label{eq:vndef}
    v_n = \frac{\int_0^{2\pi} \der \phik e^{ni \phik} \der \sigma^{e+p\to e+p+V}/\der t \der \phik \der Q^2 \der \xpom } {\int_0^{2\pi} \der \phi_{k\Delta} \der \sigma^{e+p \to e+p+V}/\der t \der \phik  \der Q^2 \der \xpom }\, .
\end{equation}

The $v_1$ and $v_2$ modulations for the exclusive \jpsi\ production at two different $\xpom$ values accessible at the EIC are shown in Fig.~\ref{fig:jpsi_vn_xdep} using the impact parameter dependent MV model initial condition coupled to the JIMWLK evolution. In case of heavy meson production, the $v_1$ modulation is found to be extremely small, as it is dominated by the off-forward phase factor whose contribution is small when the dipole size $|\rt^2| \sim 1/(Q^2+M_{\mathrm{J}/\psi}^2)$ is small. The $v_2$ modulation is significant, as it receives also a contribution from the $\cos(2\phirb)$ modulation present in the dipole-proton scattering amplitude. The modulations are suppressed at smaller $\xpom$, and this change is driven by the JIMWLK evolution. The proton grows towards small $\xpom$, which results in smaller density gradients and weaker angular dependence in the dipole scattering amplitude.

Next we compare the $v_2$ modulation we obtain using the dipole amplitude from the MV model, and from the LCPT calculation discussed briefly in Sec.~\ref{sec:setup}. We calculate $v_2$ at $\xpom=0.01$, which is not too small and the LCPT calculation, which has not been combined with a small-$x$ evolution equation, can be expected to be applicable. The results are shown in Fig.~\ref{fig:lcpt_v2}. As discussed above, the angular dependence in the ``LCPT dipole'' differs from that of the MV model, which results in $v_2$ being negative (and the magnitude being smaller) than what is obtained when the (impact parameter dependent) MV model is used. For comparison, we also calculate $v_2$ using the LCPT dipole where we have artificially removed any dependence on the dipole orientation. In this case we find a small and positive $v_2$, where the positive contribution is due to the non-zero off-forward phase $e^{-i  \vect{\delta} \cdot \vect{r}}$ e.g. in Eq.~\eqref{eq:M+-}.

\section{Conclusion}

We have shown in Ref.~\cite{Mantysaari:2020lhf} how azimuthal electron-vector meson correlations can be used to access the angular dependence of the dipole-target scattering amplitude. This, in turn, can also be related to the  gluon transversity GPD and to the gluon Wigner distribution~\cite{Hatta:2016dxp,Hatta:2017cte,Mantysaari:2019hkq}. From the Color Glass Condensate based setup, where an impact parameter dependent MV model is coupled to the JIMWLK small-$x$ evolution equation, we predict  significant ($\sim 5\%$) azimuthal modulations for exclusive \jpsi\ production in the  electron-proton collisions at the EIC. Much larger modulations are predicted for DVCS, but the Bethe-Heitler contribution may render this measurements more difficult to interpret.

In this Contribution we have studied the model dependence of the predicted azimuthal modulations, by using a dipole amplitude recently obtained as a result of explicit light cone perturbation theory calculations~\cite{Dumitru:2021tvw} (also presented in this Conference). The dependence on the dipole orientation is very different in these calculations that approach the problem from the dilute valence quark region, compare to the MV model which consider the opposite (dense) limit. We find very different predictions for the $v_2$ modulations in exclusive \jpsi\ production, even with opposite signs when the dipole amplitude from the MV model or from the LCPT calculation is used. In the future, it will be interesting to study how the transition from a dilute (described by a LCPT calculation combined with a small-$x$ evolution) to the dense region, where the MV model can be taken to be a more realistic description, takes place.


\paragraph{Funding information}
This work was supported by the Academy of Finland, projects 314764
(H.M) and 1322507 (R.P). H.M.\ is supported under the European Union’s
Horizon 2020 research and innovation programme  STRONG-2020 project (grant agreement no.
824093), and R.P.\  by the European Research Council grant agreement no. 725369. A.D.\ thanks the 
US Department of Energy, Office of Nuclear Physics, for support
via Grant DE-SC0002307. F.S. and B.P.S. are supported under DOE Contract No.~DE-SC0012704. F.S and K.R. were also supported by the joint Brookhaven National Laboratory-Stony Brook University Center for Frontiers in Nuclear Science (CFNS).
The content of this article does not reflect the official
opinion of the European Union and responsibility for the information
and views expressed therein lies entirely with the authors.
Computing resources from CSC – IT Center for Science in Espoo, Finland and from the Finnish Grid and Cloud Infrastructure (persistent identifier \texttt{urn:nbn:fi:research-infras-2016072533}) were used.



\bibliography{refs.bib}

\begin{thebibliography}{10}
\providecommand{\url}[1]{\texttt{#1}}
\providecommand{\urlprefix}{URL }
\expandafter\ifx\csname urlstyle\endcsname\relax
  \providecommand{\doi}[1]{doi:\discretionary{}{}{}#1}\else
  \providecommand{\doi}{doi:\discretionary{}{}{}\begingroup
  \urlstyle{rm}\Url}\fi
\providecommand{\eprint}[2][]{\url{#2}}

\bibitem{Ryskin:1992ui}
M.~G. Ryskin,
\newblock \emph{{Diffractive $\mathrm{J}/\psi$ electroproduction in LLA QCD}},
\newblock Z. Phys. C \textbf{57}, 89 (1993),
\newblock \doi{10.1007/BF01555742}.

\bibitem{Mantysaari:2016ykx}
H.~M\"antysaari and B.~Schenke,
\newblock \emph{{Evidence of strong proton shape fluctuations from incoherent
  diffraction}},
\newblock Phys. Rev. Lett. \textbf{117}(5), 052301 (2016),
\newblock \doi{10.1103/PhysRevLett.117.052301},
\newblock \eprint{1603.04349}.

\bibitem{Mantysaari:2016jaz}
H.~M\"antysaari and B.~Schenke,
\newblock \emph{{Revealing proton shape fluctuations with incoherent
  diffraction at high energy}},
\newblock Phys. Rev. D \textbf{94}(3), 034042 (2016),
\newblock \doi{10.1103/PhysRevD.94.034042},
\newblock \eprint{1607.01711}.

\bibitem{Mantysaari:2020axf}
H.~M\"antysaari,
\newblock \emph{{Review of proton and nuclear shape fluctuations at high
  energy}},
\newblock Rept. Prog. Phys. \textbf{83}(8), 082201 (2020),
\newblock \doi{10.1088/1361-6633/aba347},
\newblock \eprint{2001.10705}.

\bibitem{Mantysaari:2020lhf}
H.~M\"antysaari, K.~Roy, F.~Salazar and B.~Schenke,
\newblock \emph{{Gluon imaging using azimuthal correlations in diffractive
  scattering at the Electron-Ion Collider}}  (2020),
\newblock \eprint{2011.02464}.

\bibitem{Hatta:2017cte}
Y.~Hatta, B.-W. Xiao and F.~Yuan,
\newblock \emph{{Gluon Tomography from Deeply Virtual Compton Scattering at
  Small-x}},
\newblock Phys. Rev. D \textbf{95}(11), 114026 (2017),
\newblock \doi{10.1103/PhysRevD.95.114026},
\newblock \eprint{1703.02085}.

\bibitem{McLerran:1993ni}
L.~D. McLerran and R.~Venugopalan,
\newblock \emph{{Computing quark and gluon distribution functions for very
  large nuclei}},
\newblock Phys. Rev. D \textbf{49}, 2233 (1994),
\newblock \doi{10.1103/PhysRevD.49.2233},
\newblock \eprint{hep-ph/9309289}.

\bibitem{McLerran:1993ka}
L.~D. McLerran and R.~Venugopalan,
\newblock \emph{{Gluon distribution functions for very large nuclei at small
  transverse momentum}},
\newblock Phys. Rev. D \textbf{49}, 3352 (1994),
\newblock \doi{10.1103/PhysRevD.49.3352},
\newblock \eprint{hep-ph/9311205}.

\bibitem{Mueller:2001uk}
A.~H. Mueller,
\newblock \emph{{A Simple derivation of the JIMWLK equation}},
\newblock Phys. Lett. B \textbf{523}, 243 (2001),
\newblock \doi{10.1016/S0370-2693(01)01343-0},
\newblock \eprint{hep-ph/0110169}.

\bibitem{Mantysaari:2019hkq}
H.~M\"antysaari, N.~Mueller, F.~Salazar and B.~Schenke,
\newblock \emph{{Multigluon Correlations and Evidence of Saturation from Dijet
  Measurements at an Electron-Ion Collider}},
\newblock Phys. Rev. Lett. \textbf{124}(11), 112301 (2020),
\newblock \doi{10.1103/PhysRevLett.124.112301},
\newblock \eprint{1912.05586}.

\bibitem{Iancu:2017fzn}
E.~Iancu and A.~H. Rezaeian,
\newblock \emph{{Elliptic flow from color-dipole orientation in pp and pA
  collisions}},
\newblock Phys. Rev. D \textbf{95}(9), 094003 (2017),
\newblock \doi{10.1103/PhysRevD.95.094003},
\newblock \eprint{1702.03943}.

\bibitem{Dumitru:2021tvw}
A.~Dumitru, H.~M\"antysaari and R.~Paatelainen,
\newblock \emph{{Color charge correlations in the proton at NLO: beyond
  geometry based intuition}}  (2021),
\newblock \eprint{2103.11682}.

\bibitem{Dumitru:2020gla}
A.~Dumitru and R.~Paatelainen,
\newblock \emph{{Sub-femtometer scale color charge fluctuations in a proton
  made of three quarks and a gluon}},
\newblock Phys. Rev. D \textbf{103}(3), 034026 (2021),
\newblock \doi{10.1103/PhysRevD.103.034026},
\newblock \eprint{2010.11245}.

\bibitem{Hatta:2016dxp}
Y.~Hatta, B.-W. Xiao and F.~Yuan,
\newblock \emph{{Probing the Small-$x$ Gluon Tomography in Correlated Hard
  Diffractive Dijet Production in Deep Inelastic Scattering}},
\newblock Phys. Rev. Lett. \textbf{116}(20), 202301 (2016),
\newblock \doi{10.1103/PhysRevLett.116.202301},
\newblock \eprint{1601.01585}.

\end{thebibliography}

\nolinenumbers

\end{document}